\documentclass[twocolumn,showpacs,preprintnumbers,amsmath,floatfix,amssymb]{revtex4}

\usepackage{graphicx}
\usepackage{dcolumn}
\usepackage{bm}

\begin{document}


\title{Monte Carlo Simulation of Small-World Models with Weak
  Long-Range Interactions}

\author{X.~Zhang and M.~A.~Novotny}
\affiliation{Dept.\ of Physics and Astronomy;
Center for Computational Sciences;
P.O.\ Box 5167;
Mississippi State University;
Mississippi State, MS 39762-5167}
\email{man40@ra.MsState.edu}
\homepage{http://www.msstate.edu/dept/physics/profs/novotny.html}

\date{\today}

\begin{abstract}
Large-scale Monte Carlo simulations, together with scaling, are used to
obtain the critical behavior of the Hastings long-range model and two
corresponding models based on small-world networks. These models have
combined short- and long-range interactions. For all three models, the
critical scaling behavior predicted by Hastings [Phys. Rev. Lett. 91,
098701] is verified. 
\end{abstract}

\pacs{64.60.-i,64.60.Fr,89.75.Hc}
\maketitle
\section{Introduction}

Traditionally, studies of materials have concentrated on either models
based on regular graphs or random graphs (such as the models on percolating lattices \cite{KIRK73}, the $z$ model
\cite{SCAL91, NOVO05c}, or the model in Ref. \cite{RAME04}). Recently,
much more attention has been paid to networks that are midway between random graphs
and regular graphs, for example, scale-free networks
\cite{IGLO02}. The critical behavior, as well as the transport
properties, are different from that of the same model on a regular
graph. The small world (SW) graph \cite{BARA02} is a type of graph
that interpolates between a regular graph and a random graph. These
graphs have been widely used in many areas, for example, in parallel
simulations \cite{KORN03,KORN99,KORN00,KOLA03,KORN02} and in studies
of the Internet \cite{BARA02}. Since a particular material is related to a
given graph with interactions based on the two-body interaction
approximation \cite{CHAR85}, many models of materials have been
studied on small world graphs. These models include the Ising model,
Heisenberg model, and random-walker models
\cite{GITT00,BARR00,PEKA01,HONG02b,KIM01,HONG02,HERR02,JEON03,HAST03,KOZM04,JEON05,KOZM05,NOVO04,NOVO05,NOVO05b,NOVO05c}.
Such studies show that these models exhibit mean-field behavior.

It is not easy to analyze small-world models theoretically because of
the randomness of the long-range bonds \cite{BARA02}. Hastings introduced a
long-range model to explain the critical phenomena in small-world
systems analytically \cite{HAST03}. His model, in which long-range
interactions are distributed to all the spins in the system, has no
randomness and is easier to analyze. He claimed that his model has the
same universal behavior as the small world model, with some
constrains, and presented a criterion for the crossover to the
underlying mean-field behavior. Thus far, there are no experimental or
computer simulation verifications of Hastings long-range model, as
well as its comparision with other small-world models. In this
article, Monte Carlo simulations for the $2$-dimensional Hastings
model and two corresponding SW models with Ising spins are presented and the
critical behavior is analyzed. 

\section{Model and Methods}

All three models studied start with a $2$-dimensional $L \times L$
square lattice with periodic boundary conditions and an Ising spin
with $s_i \pm 1$ located on each site, and a nearest neighbor
interaction constant $J$. For the first model, the small-world Ising
model, randomly chosen pairs of Ising spins are connected with a bond,
a fixed number ($w$) of these small-world bonds are added. Once a SW bond is added to a pair of spins, no other SW bond is
allowed to be added to these spins. These SW bonds do not change
during the simulation, i.e. they are quenched. So the total number of
Ising spins in the model is $N = L^2$, and the number of SW bonds is
less than $\frac{L^2}{2}$. Thus, the ratio of the number of
small-world bonds to the regular nearest neighbor bonds is
$p=\frac{w}{2L^2}$. On average, a spin in the system has $z=4+
\frac{2w}{L^2}=4+4p$ nearest neighbor bonds. We suppose the
interaction via SW bonds is the same strength as that via regular
lattice bonds. The Ising Hamiltonian of such a system is given by 
\begin{equation}
\label{eqHAMSW}
{\cal H} = - J \sum_{\langle i,j\rangle} s_is_{j}
- J \sum_{\rm SW} s_i s_j
\>.
\end{equation}
The first sum is over the four nearest neighbor spins on the square
lattice, and the second sum is over the $w$ SW bonds. We use $J=1$ in
this article. 

Hastings constructed his long-range model by giving each spin on the
lattice a weak coupling, of order $\frac{p}{N}=\frac{p}{L^2}$, to
every other spin in the lattice, instead of adding long-range bonds
with probability $p$ \cite{HAST03}. Hence the Hastings' Hamiltonian
can be written as 
\begin{eqnarray}
{\cal H} &=& - J \sum_{\langle i,j\rangle} s_is_j
- J \frac{p}{2N} \sum_{i, j \neq i} s_i s_j
\nonumber\\
&=& - J \sum_{\langle i,j\rangle} s_i s_j
- J \frac{p}{2N} \left( N \sum_{i}  M s_i - N \right). \>
\label{eqHAMHAST}
\end{eqnarray}
Here $M$ is the magnetization density,
\begin{equation}
\label{eqM}
M = \frac{1}{N} \sum_{i} s_i
\>.
\end{equation}
We utilized standard Monte Carlo (MC) simulations \cite{LAND00}. A
Glauber flip probability with the site for an attempted update chosen
at random is used. In our implementation, two random numbers, $r_1,
r_2$, are generated. The first number is used to randomly choose a
spin for the spin flip attempt, the second is used to determine whether the
chosen spin should be flipped. If
\begin{equation}
r_2 \leq
\frac{\exp^{-E_{\rm new}/k_{\rm B}T}}
{\exp^{-E_{\rm old}/k_{\rm B}T}+\exp^{-E_{\rm new}/k_{\rm B}T}}
\>,
\label{eq:Glauber}
\end{equation}
the chosen spin will be flipped. Here $T$ is the temperature and
 $k_{\rm B}$ is Boltzmann's constant (in our units $k_{\rm
 B}$$=$$1$). The current energy is $E_{\rm old}$ and $E_{\rm new}$ is
 the energy if the chosen spin is flipped.

We also study another small-world model, one with annealed SW
 bonds. The model is on the square lattice, with a long-range
 interaction the same as the regular lattice interaction $J$. At each
 Monte Carlo spin flip trial, randomly choose a spin and for this one
 update attempt, add with probability $p$ a small-world connection
 between this spin and another randomly chosen spin. Calculate the
 energies $E_{\rm old}$ and $E_{\rm new}$ using the four
 square-lattice nearest neighbor spins and the small-world connection
 (if it is added), with use of Eq.~(\ref{eq:Glauber}), flip the chosen
 spin. This Monte Carlo process is similar to the spin-exchange
 process used by R\'{a}cz et al \cite{RACZ90, RACZ91}. Since the
 long-range random connection is annealed in this model, we call this
 model the annealed small-world model.

The Monte Carlo simulation is performed on a parallel computer with
each processing element running a particular temperature for systems
with size less than $N=256^2$, while up to four processing elements
running a particular temperature for the largest system size
$N=384^2$. The SPRNG \cite{SPRNG} random number generator is used. A 
number of quantities are measured, but only the order parameter
($|M|$) and the Binder $4$th order cumulant ($U_4$) are reported here,
$|M| = \frac{1}{N K} \sum_{j=1}^K\left | \sum_{i=1}^N s_i \right |_j$
and $U_4=1- \frac{\left\langle M^4\right\rangle}{3 \left\langle
    M^2\right\rangle^2}$. The summation index $j$ runs over the $K$
different configurations generated in the Monte Carlo simulation. Up
to 128 processing elements were used in our simulations. It took about
$40$ hours per data run for system size $N=256^2$ and $70$ hours for
system size $N=384^2$. For each temperature, averages are taken using
$K$$=$$10^6$ Monte Carlo Steps per Spin (MCSS).
 
Hastings developed a general scaling method by using renormalization
group theory combined with mean-field techniques for small world
systems. He predicted that for small-world systems with small $p$, the
critical point shifts away from the Ising critical point, from
Eq. (9) in \cite{HAST03},
\begin{equation}
\label{eq:weakcrossTctilde}
{\widetilde{T}}_c - T_c = A_t p^{1/\gamma}
\>.
\end{equation}
Here $\gamma$ is the critical exponent for the susceptibility for the local regular lattice system, $\widetilde{T}_c$ is the critical temperature of SW system, and $T_c$ is the critical temperature of the local system. Here the local system
is a $2d$ square Ising lattice, so $T_c=2.2691 \cdots $. The prefactor
$A_t= (A_{\chi}^{+})^{1/\gamma}$, where $A_{\chi}^{+}$ is the
prefactor in the scaling of the susceptibility $\chi$. It is also
predicted that below the critical point, $ T < \widetilde{T}_c$, the
magnetization in the limit $N \rightarrow \infty$ is (from Eq. (10) in
\cite{HAST03})
\begin{equation}
\label{eq:weakcrossMtctildetc}
|M| = A_M {\widetilde{T}}_c^{\frac{1}{2}}
\left( 1 - \frac{T}{{\widetilde{T}}_c} \right)^{\frac{1}{2}}
\left( {\widetilde{T}}_c - T_c \right)^{\beta - \frac{1}{2}}
\>.
\end{equation}
Here $A_M$ is a prefactor and $\beta$ is the critical exponent for the
order parameter for the local regular lattice system. In the models
studied in this article, the local system is a $2d$ Ising model, so
$\beta = \frac{1}{8}$ and $\gamma = \frac{7}{4}$. Hence $|M| = {\widetilde A}
\sqrt{\widetilde{T}_{\rm c}-T}$ for $T < \widetilde{T}_{\rm c}$, and
the mean field amplitude $\widetilde A \propto
p^{\frac{\beta-\frac{1}{2}}{\gamma}}$ diverges as the long-range
interaction strength $p$ (or the ratio of SW bonds to regular lattice
bonds $p$) approaches zero. The mean field region for this scaling is
small \cite{HAST03}, with mean field behavior only over a region given by
\begin{equation}
|T_{\rm c}-T| \propto p^{\frac{1}{\gamma}} \propto
|T_{\rm c}-{\widetilde{T} }_{\rm c}|
\>.
\label{eq:MFclose}
\end{equation}
Note that this type of scaling is related to the coherent anomaly
method pioneered by Suzuki \cite{MSUZ95}. Also, this scaling can only
be expected to be seen in systems with a small amount of long-range
interactions ($p$ small). For SW systems with a large amount of
long-range bonds ($p=0.25$), even if the long-range interaction is
weak ($0.01J, \> 0.05J, \> 0.1J, \> 0.5J$), this predicted scaling was
not observed \cite{NOVO05c} for the system sizes we studied.

For small $p$, as the system size increases, the crossing point of
$U_4$ approaches the point where $U_4$ is the predicted value for
infinite size mean field models, $U_{4,\infty}^{\rm (MF)}=0.2705
\cdots$ \cite{NOVO05c, BREZ85}. We use the temperature where $U_4 =
U_{4,\infty}^{\rm (MF)}$ as the estimate for the critical temperature
$\widetilde{T}_{\rm c}$ in our scaling.

\section{Results and Scaling}
\label{Sec:RS}

\begin{figure}
\includegraphics[width=.8500\columnwidth]{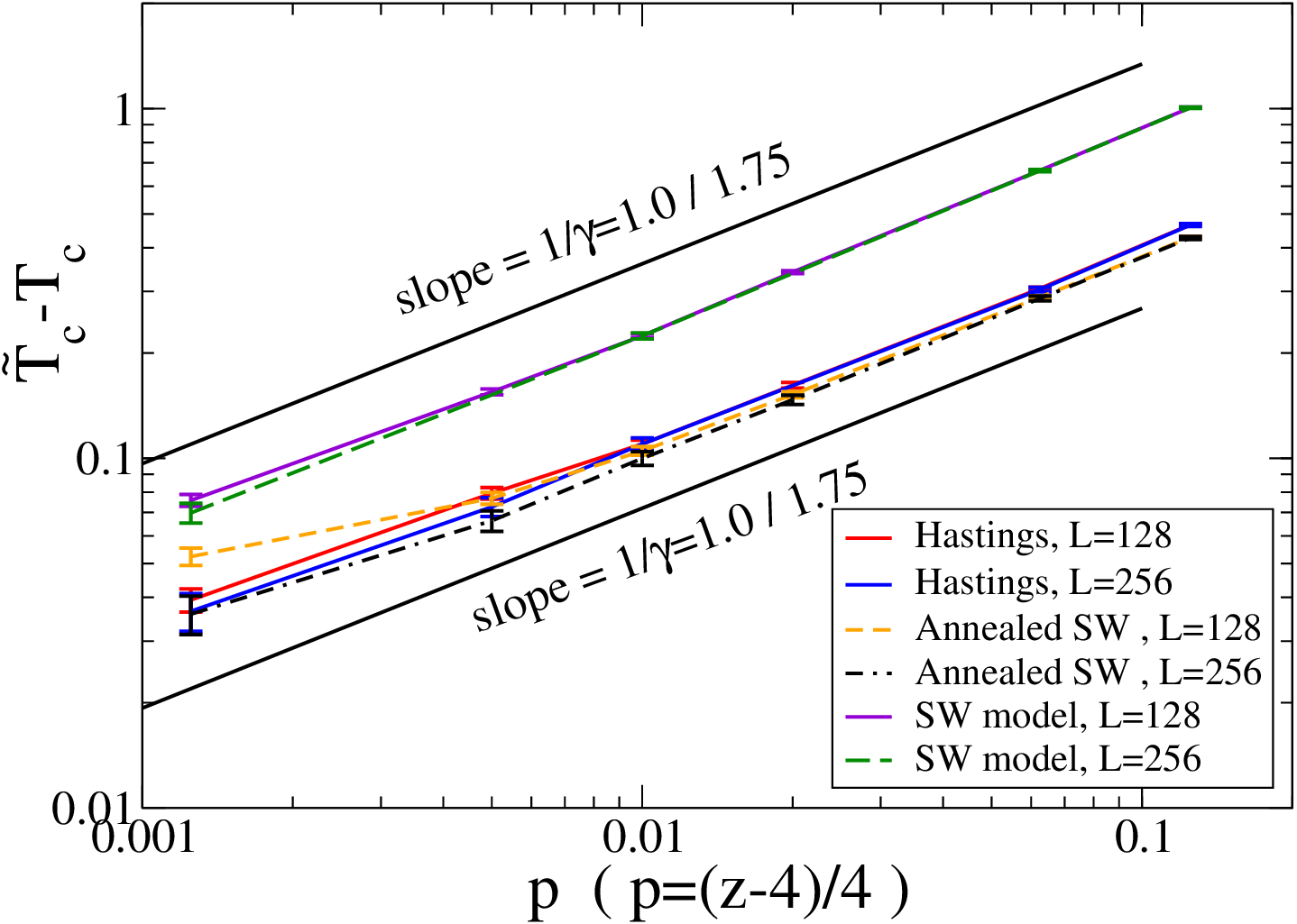}
\caption[]{(color online) The critical temperature relation from
  Eq.~(\ref{eq:weakcrossTctilde}). The critical temperature of the
  Hastings model is much smaller than that of the corresponding $2d$
  SW model (upper curves), but very close to that of the annealed SW
  model. The data show that Hastings' prediction about the critical
  temperature works well.}
\label{xzh_fig1}
\end{figure}

\begin{figure}
\vspace{0.2in}
\includegraphics[width=.8500\columnwidth]{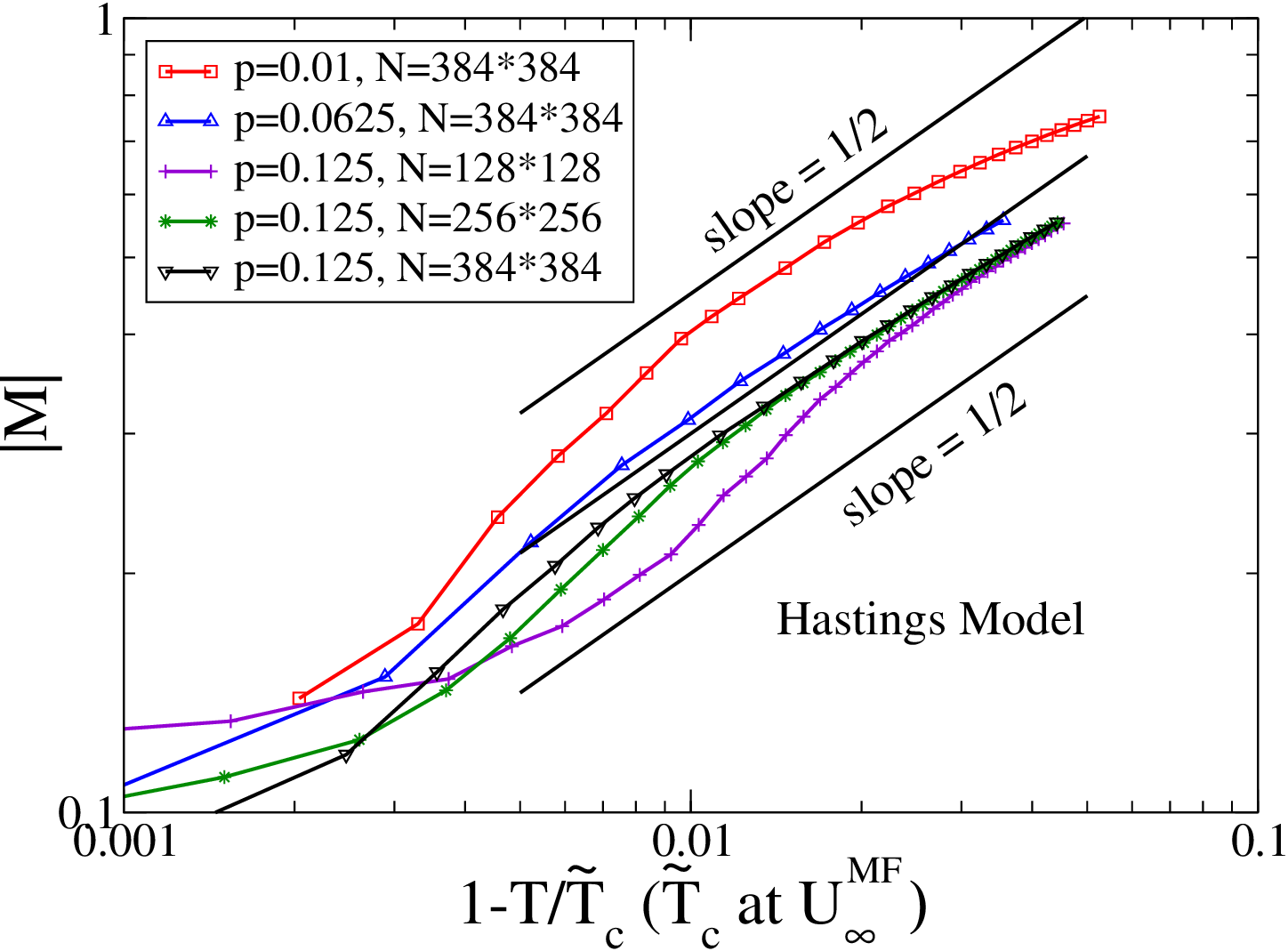}
\caption[]{(color online) The order parameter $|M|$
  vs. $t=1-\frac{T}{\widetilde{T}_{\rm c}}$ for the Hastings
  long-range model.}
\label{xzh_fig2}
\end{figure}

Simulations were performed for $p=0.00125, \> 0.005, \> 0.01, \> 0.0625$ and
$0.125$ for the Hastings model and the annealed small-world model. For
the $2d$ SW model, simulations were performed for $z=4.005, \> 4.02, \> 4.04,
\> 4.254$ and $4.5$, which correspond to the above $p$ values. The
results show that the critical temperature for the $2d$ SW model is
larger than that for the Hastings model for the same $p$. That is what
is expected since the spreading of pair-wise long-range interactions
to every spin in the system weakens the long-range effect. The
critical temperature for the annealed SW model is a little bit smaller
than that of the Hastings
model. Fig.~(\ref{xzh_fig1}) shows the critical temperature relation
from Eq.~(\ref{eq:weakcrossTctilde}) for all these three models, as well
as the theoretical prediction. For each model, the data falls on a
straight line of the Hastings predicted slope of $\frac{1}{\gamma}$. The
small deviation from the slope of $\frac{1}{\gamma}$ for the smaller
values of $p$ is due to fluctuations and statistical
uncertainty. For the Hastings long-range model, the error estimate from
$16$ independent samples for $p=0.00125$ is $\sigma=0.0030$ for
$N=128^2$ and $\sigma=0.0045$ for $N=256^2$. The error estimates for the other
systems is of the same order.

\begin{figure}
\includegraphics[width=.8500\columnwidth]{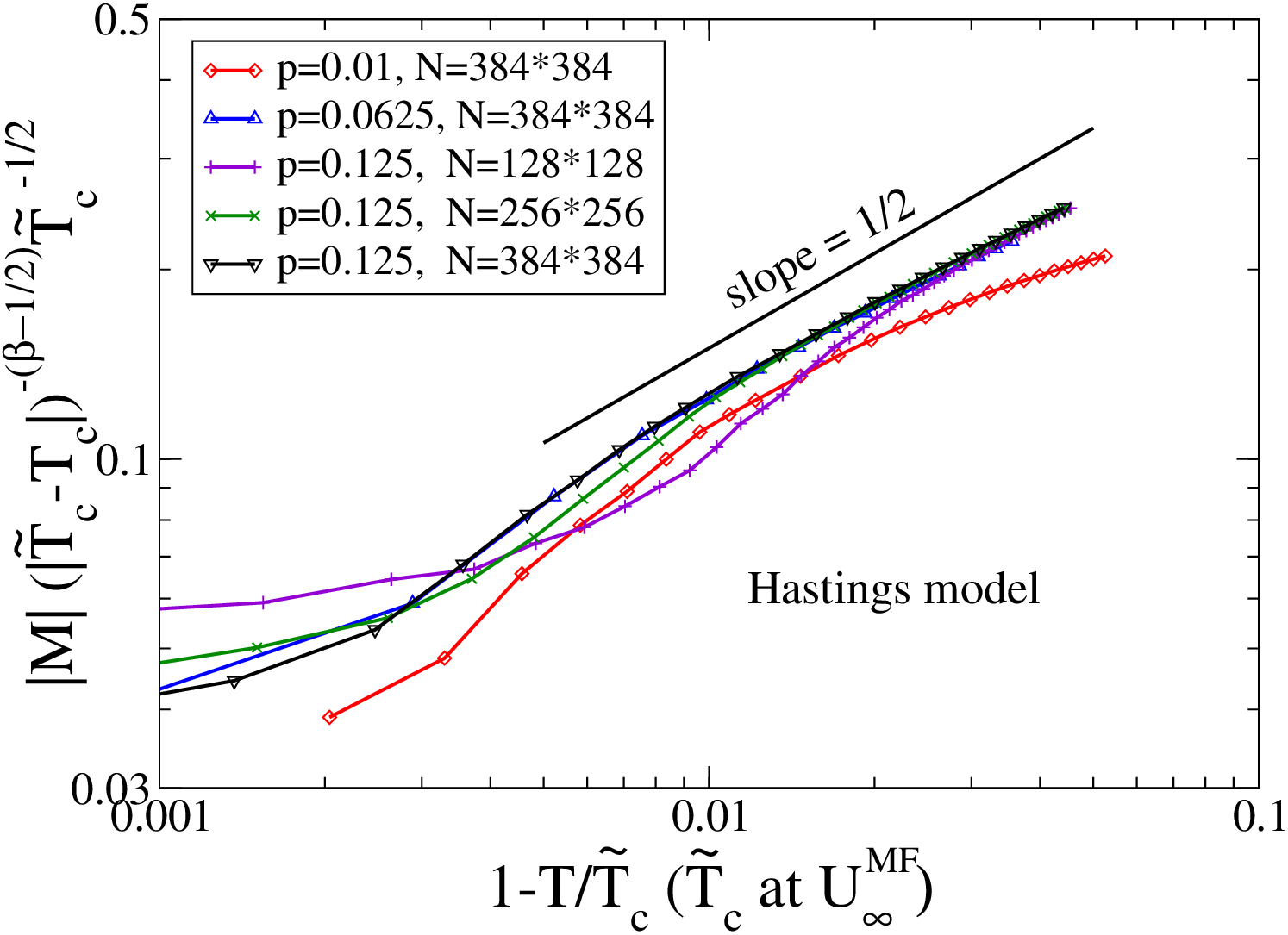}
\caption[]
{(color online) The order parameter scaling from
  Eq.~(\ref{eq:weakcrossMtctildetc}) for the Hastings long-range model.}
\label{xzh_fig3}
\end{figure}

Fig.~(\ref{xzh_fig2}) is the simulation result for the order parameter
vs. the reduced temperature $t$, where $t = 1 -
\frac{T}{\widetilde{T}_{\rm c}}$, for the Hastings model. Only the
branches for $T < \widetilde{T}_{\rm c}$ are shown. It can be seen from the
figure, the data curve for $N=384^2$ and $p=0.125$ approaches a slope
$\frac{1}{2}$ line in the region near $t < 0.04$, which is consistant with
the estimate from Eq.~(\ref{eq:MFclose}) (for $p=0.125$, and 
the largest system size we studied, $N=384^2$, $\widetilde{T}_{\rm c} =
2.734$). The deviation from the slope of $\frac{1}{2}$ at small $t$ is
due to finite size effects. This is seen by comparing systems for $p=0.125$,
the smaller lattices ($N=128^2$ and $256^2$) deviate from the
$N=384^2$ data at small $t$. For a fixed value of $N$, the region with
the expected slope of $\frac{1}{2}$ is also not seen for values of $p$
that are too small (as for $N=384^2$ and $p=0.01$). Consequently, the
prediction of Hastings \cite{HAST03} with a slope of $\frac{1}{2}$ is
only seen for large enough values of $N$ and values of $p$ which are
small (but not too small for a fixed $N$).

\begin{figure}
\includegraphics[width=.8500\columnwidth]{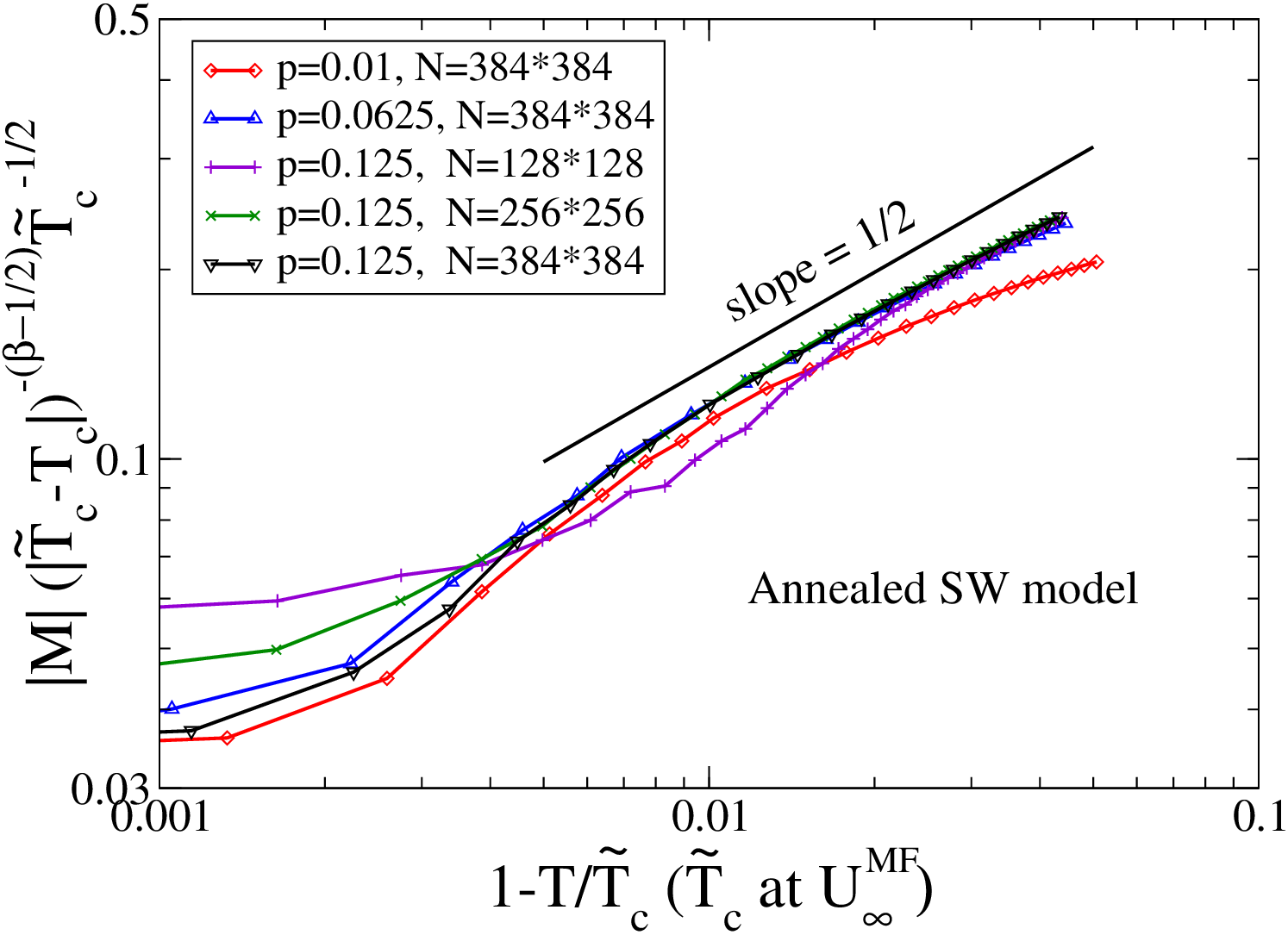}
\caption[]
{(color online) The order parameter scaling from
  Eq.~(\ref{eq:weakcrossMtctildetc}) for the annealed SW model.}
\label{xzh_fig4}
\end{figure}

\begin{figure}
\vspace{0.2in}
\includegraphics[width=.8500\columnwidth]{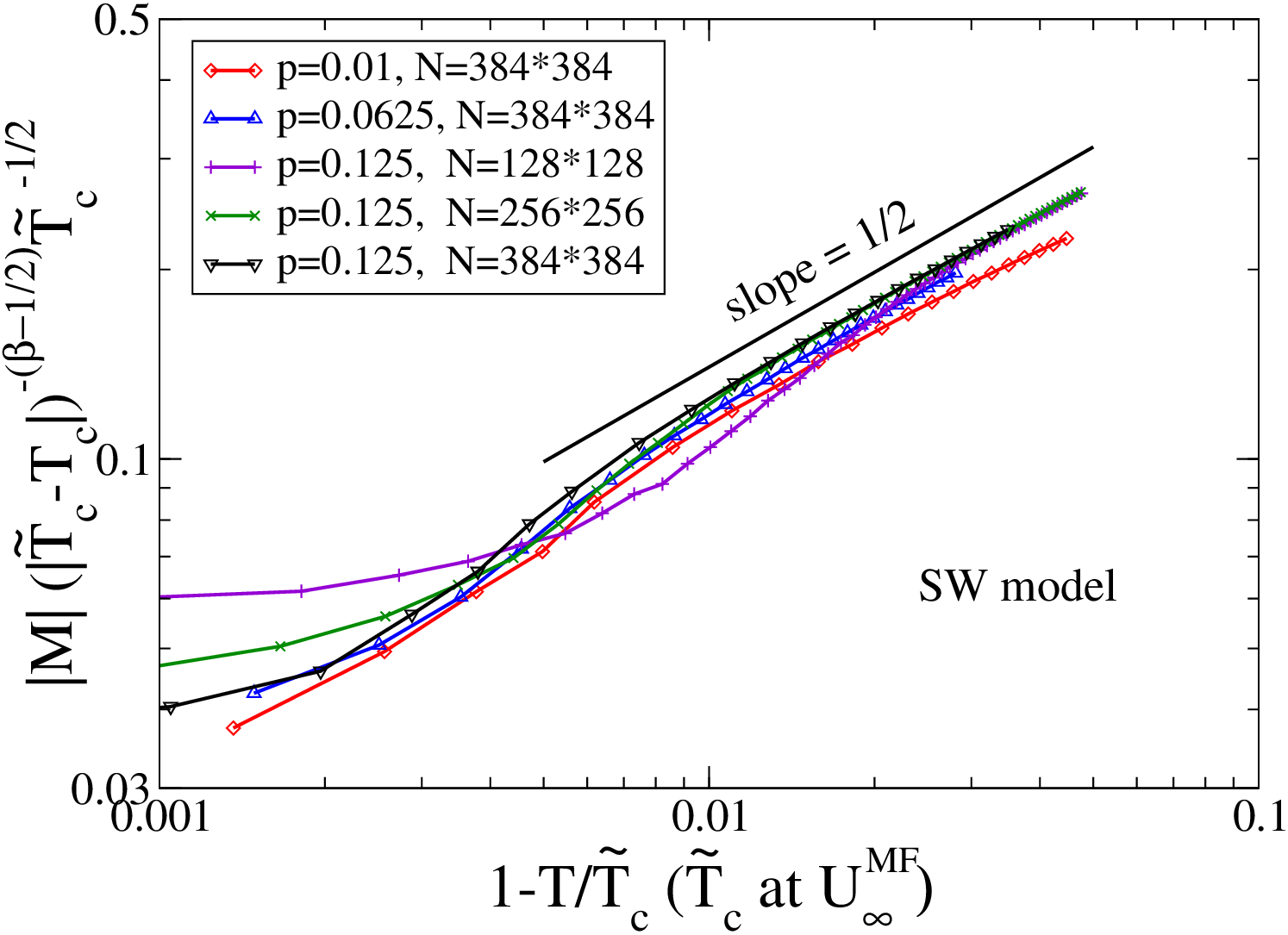}
\caption[]
{(color online) The order parameter scaling from
  Eq.~(\ref{eq:weakcrossMtctildetc}) for the $2d$ SW model.}
\label{xzh_fig5}
\end{figure}

Fig.~(\ref{xzh_fig3}) to Fig.~(\ref{xzh_fig5}) are scaling results for the
order parameter $|M|$ using Eq.~(\ref{eq:weakcrossMtctildetc}) for the
Hastings model, the annealed SW model, and the SW model. From
Eq.~(\ref{eq:MFclose}), the mean field critical region is proportional
to $\widetilde{T}_{\rm c}-T_{\rm c}$. The smaller the $p$ value, the
smaller is $\widetilde{T}_{\rm c}-T_{\rm c}$ and the narrower is the
mean field critical region \cite{HAST03}. In the mean field region,
the scaled data points collapse to a line of slope $\frac{1}{2}$. For
the Hastings model, shown in Fig.~(\ref{xzh_fig3}), the scaled data
points fall on a region close to a line of slope $\frac{1}{2}$ near
$t=1-\frac{T}{\widetilde{T}_{\rm c}}=0.01$ for all $p$ for larger $N$, and
expands beyond $t=0.02$ on the upper side and $0.008$ on the lower
side for values of $p$ that are larger ($p=0.0625$ and $0.125$). The
annealed SW model and the SW model present the same behavior,
Fig.~(\ref{xzh_fig4}) and Fig.~(\ref{xzh_fig5}). It is expected that,
for a fixed size system, the mean field region will be wider as $p$
increases (but is still small enough so that the long-range
interaction is weak). The scaling result for the $2d$ SW model illustrates
this. In Fig.~(\ref{xzh_fig5}), for system size $N=384^2$, the
mean field region of the largest value of $p=0.125$ expands much
further to smaller $t$ than that of $p=0.0625$ and $p=0.01$. For a
fixed $p$, the mean field region is wider for larger size systems. In
Fig.~(\ref{xzh_fig3}), Fig.~(\ref{xzh_fig4}) and
Fig.~(\ref{xzh_fig5}), for $p=0.125$, as the 
system size increases from $N=128^2, 256^2$ to $N=384^2$, the mean
field region extends to smaller $t$, or, to the area where the
temperature $T$ is very close to the critical temperature
$\widetilde{T}_{\rm c}$. But for much smaller values of $p$, for
example, $p=0.01$, the mean field region is hardly seen even for
$N=384^2$. Simulations for larger size systems are needed to penetrate
the mean field region as Hastings predicted \cite{HAST03} for much
smaller $p$. Unfortunately, we can not run larger size systems due
to computer limits.  

From the above analysis, the mean field region predicted by Hastings
\cite{HAST03} for the order parameter is seen for large enough
values of $N$ and values of $p$ which are small (but not too small for
a fixed $N$) in all three Ising models we studied, the Hastings
model, the annealed SW model and the SW model. It is hard to say for
which model Eq.~(\ref{eq:weakcrossMtctildetc}) works better. Although
for $p=0.125$ and $N=384^2$, these models present a reasonable scaling
result, the scaling result for a smaller $p$, here $p=0.01$, is not
good and the mean field region is not obvious for our system sizes and
statistics. It can not be determined whether this deviation from the predicted
scaling is due to fluctuations or to finite size effects or some combination.

\section{Discussion and Conclusions}

Starting from the square lattice Ising model, we constructed a
 small-world Ising model and an annealed small-world model that have the same
  critical behavior as the Hastings long-range model. These two Ising
 models have been simulated, as well as the Hastings model. Hastings'
 predictions for the critical temperature and the order parameter
 scaling \cite{HAST03}  are verified. Results show the predicted
 scaling relation in \cite{HAST03} for the critical point
 $\widetilde{T}_{\rm c}$, Eq.~(\ref{eq:weakcrossTctilde}), works well
 for all three models, Fig.~(\ref{xzh_fig1}). The mean field behavior
 predicted by Hastings \cite{HAST03},
 Eq.~(\ref{eq:weakcrossMtctildetc}), is seen in the mean field region,
 Eq.~(\ref{eq:MFclose}), for system sizes $N=256^2, N=384^2$ and not too
 small $p$ ($p=0.0625$ and $0.125$), Fig.~(\ref{xzh_fig3}),
 Fig.~(\ref{xzh_fig4}) and Fig.~(\ref{xzh_fig5}). To see whether
 Hastings' predictions work for much smaller $p$ (for example, $p \sim
 0.01$), simulations for much larger size systems ($N >> 384^2$) and
 much higher statistics would be required.

\section{Acknowledgements}

We acknowledge useful discussions with a number of people, particularly
Gyorgy Korniss, Per Arne Rikvold, and Zoltan Toroczkai. Supported in
part by NSF grants DMR-0120310, DMR-0113049, DMR-0426488, and DMR-0444051.
Computer time from the Mississippi State University High Performance
Computing Collaboratory (HPC$^2$) was critical to this study.

\end{document}